\documentclass[pre,amssymb,reprint,groupaddress,superscriptaddress,showpacs,longbibliography]{revtex4-1}
\usepackage{amsmath}
\usepackage{amsfonts}
\usepackage{amssymb}
\usepackage{epsfig}
\usepackage{graphicx}
\usepackage{color}
\definecolor{darkblue}{rgb}{0,0,0.6}
\usepackage[colorlinks,linkcolor=darkblue,citecolor=darkblue,urlcolor=darkblue]{hyperref}

\newcommand{\beq}{\begin{equation}} \newcommand{\eeq}{\end{equation}}
\newcommand{\bal}{\begin{align}} \newcommand{\eal}{\end{align}}
\newcommand{\fig}[1]{Fig.~\ref{#1}}

\newcommand{\Eq}[1]{Equation~(\ref{#1})}
\definecolor{teal}{rgb}{0.0,0.5664,0.5742}
\definecolor{strawberry}{rgb}{1.0,0.0,0.5}

\newcommand{\tD}{\tilde{D}}
\newcommand{\tom}{\tilde{\omega}}

\begin{document}

\title{Dynamics of viscous liquids and the Random Barrier Model}

\author{Thomas B. Schr{\o}der}
\email{tbs@ruc.dk}
\affiliation{``Glass and Time,'' IMFUFA, Department of Science and Environment, Roskilde University, P.O. Box 260, DK-4000 Roskilde, Denmark}

\author{Jeppe C. Dyre}
\email{dyre@ruc.dk}
\affiliation{``Glass and Time,'' IMFUFA, Department of Science and Environment, Roskilde University, P.O. Box 260, DK-4000 Roskilde, Denmark}

\author{Camille Scalliet}
\email{camille.scalliet@ens.fr}
\affiliation{Laboratoire de Physique de l'Ecole Normale Sup\'erieure, ENS, Universit\'e PSL, CNRS, Sorbonne Universit\'e, Universit\'e de Paris, F-75005 Paris, France}

\date{\today}

\begin{abstract}
This paper combines the particle-swap Monte Carlo algorithm with long GPU molecular dynamics simulations to analyze the dynamics of a ternary Lennard-Jones glass-forming liquid in the extremely viscous regime. The focus is on the inherent dynamics, obtained by quenching configurations along the configuration-space trajectory into their inherent state. We compare how two functional forms, the von Schweidler law and the random barrier model (RBM) prediction in the extreme disorder limit, fit data for the inherent mean-squared displacement as a function of time. We find that the RBM, which has no dimensionless free parameters, generally fits the data better than the von Schweidler law, despite the latter's one dimensionless free parameter. In particular, this implies that the RBM predicts the value of the diffusion coefficient from short-time simulation data more accurately than does the von Schweidler expression. It remains an open question why the RBM reproduces well the inherent data despite this model's (unrealistic) assumption of identical energy minima.
\end{abstract}

\maketitle

\section{\label{sec:introduction}Introduction}

Experimental works of Gainaru and coworkers highlighted a possible universality of the complex frequency-dependent fluidity of non-polymeric viscous liquids (the inverse of the shear viscosity) \cite{gainaruprl2017,gainarupre2019}. For nine quite different glass-formers -- including van der Waals, ionic, and hydrogen-bonding liquids -- the real part of the fluidity was found to be very similar as a function of frequency. This ``universal'' shape was found to be well described by the extreme disorder limit of the random barrier model (RBM). Subsequent simulations showed that the RBM fits also the mean-squared displacement (MSD) of a simple binary glass-forming liquid~\cite{solidlike}. In particular, the so-called \emph{inherent} MSD \cite{sch00}, where the MSD short-time plateau deriving from thermal vibrations is removed, was found also to be well fitted by the RBM prediction, which has no free shape parameters.   

The RBM was proposed as an idealized model of AC conduction in amorphous solids \cite{sch00,dyr88}. The model considers independent particles jumping between nearest-neighbor same-energy sites of a cubic lattice connected by energy barriers drawn from some probability distribution. The model was later solved in the extreme disorder limit of an effective-medium approximation, i.e., at low temperatures where the jump rates cover many decades, and the analytical solution agrees very well with numerical results \cite{sch08}.

The experimental and simulation results described above suggest universality in the motion of atoms or molecules in highly viscous liquids. This raises two questions: i) The picture of quenched disorder with identical site energies, as in the RBM, is not how one usually think about liquids, neither in 3d space, nor in the high-dimensional configuration space. To the contrary, viscous liquids are well known to have a broad distribution of inherent state energies (potential-energy minima) leading to a higher specific heat in the liquid than in the glass phase. So, why does the RBM describe viscous liquid dynamics well? ii) The dynamics of viscous liquids has been studied for many years, both in experiments and in simulations; why has the apparent RBM universality not been observed before?

In this article we present simulations of a ternary glass former carried out in order to investigate these questions further. Thanks to the addition of a third particle type compared to the model investigated in Ref.~\onlinecite{solidlike}, the liquid can be equilibrated efficiently down to lower temperatures. We test the agreement with the RBM by performing simulations down to temperatures at which the dynamics is roughly 30 times slower than those investigated in Ref.~\onlinecite{solidlike}. This is done by combining state-of-the-art numerical techniques. Specifically, we take advantage of a recent extension of the swap algorithm to a model of metallic glasses in order to generate equilibrium configurations at very low temperatures \cite{parmar_ultrastable_2020}, and this is combined with state-of-the-art graphics-processing unit (GPU) molecular-dynamics simulations that efficiently access equilibrium dynamics at low temperatures over long time scales~\cite{rumd}. 

\section{Methods}

\subsection{Model}

We study a three-dimensional model for metallic glasses, employing the ternary mixture developed in Ref.~\cite{parmar_ultrastable_2020}. This model extends the well-known Kob-Andersen (KA) model~\cite{kob_testing_1995} by adding a third particle type, which ensures resistance against crystallization and increased efficiency for equilibration with particle-swap dynamics, as detailed in Ref.~\cite{parmar_ultrastable_2020} where the model is referred to as KA$_2$.

The KA$_2$ model involves three particle types, A (large), B (small), and C (medium), in the ratio A:B:C=4:1:1. Two particles $i$ and $j$ at positions $\mathbf{r}_i$ and $\mathbf{r}_j$, separated by the distance $r = |\mathbf{r}_i - \mathbf{r}_j|$, interact via a smoothed Lennard-Jones (LJ) potential,
\begin{equation}
    v_{ij}(r) = 4 \epsilon_{ij} \left[ \left( \dfrac{\sigma_{ij}}{r} \right)^{12} - \left( \dfrac{\sigma_{ij}}{r} \right)^6 \right] + S_{ij}(r)  
\end{equation}
whenever $r<2.5 \sigma_{ij}$. The smoothing polynomial removing discontinuities resulting from the truncation is given by 
\begin{equation}
	S_{ij}(r) = 4 \epsilon_{ij} \left[ C_0 + C_2 \left( \frac{r}{\sigma_{ij}} \right)^2 + C_4 \left(\frac{r}{\sigma_{ij}} \right)^4 \right] \ .
\end{equation} 
Here $C_0 = 10/r_c^6-28/r_c^{12}$, $C_2 = 48/r_c^{14}-15/r_c^{8}$ and $C_4 = 6/r_c^{10}-21/r_c^{16}$ ensures continuity of the potential $v_{ij}(r)$ and its first two derivatives at the cutoff distance $r_c\equiv 2.5 \sigma_{ij}$. 

The unit length is $\sigma_{AA}$ and the unit energy is $\epsilon_{AA}$, henceforth denoted $\sigma$ and $\epsilon$ for simplicity. In terms of these quantities the remaining interaction parameters are $\epsilon_{AB}=1.5 \epsilon$, $\epsilon_{AC}=0.9 \epsilon$, $\epsilon_{BB}=0.5 \epsilon$, $\epsilon_{BC}=0.84 \epsilon$ and $\epsilon_{CC}=0.94 \epsilon$ for the energies; and $\sigma_{AB}=0.8 \sigma$, $\sigma_{AC}=1.25 \sigma$, $\sigma_{BB}=0.88 \sigma$, $\sigma_{BC}=1.00 \sigma$ and $\sigma_{CC}=0.75 \sigma$ for the interaction ranges. All particles have the same mass $m$. The unit time is $\tau_{LJ} = \sqrt{m \sigma^2/\epsilon}$. 

We simulated $N = 12 000 $ particles at number density $\rho=N/L^3 = 1.35$ in a cubic box of linear size $L$ with periodic boundary conditions. This particle density is higher than the commonly investigated $\rho = 1.2$~\cite{kob_testing_1995}, but the average particle size of the KA$_2$ model is slightly smaller than that of the standard KA model. Indeed, for $\rho=1.2$ the attractive potential gives rise to a liquid-gas spinodal at low temperature that may intersect the glass transition line~\cite{sastry_liquid_2000}, leading to a gas-glass instability~\cite{testard_influence_2011}. Choosing $\rho = 1.35$ ensures stability of the liquid, as well as positive values for its equilibrium pressure and the corresponding inherent states down to the lowest temperatures investigated ($T=0.509$).

The temperature scales relevant for this model at $\rho = 1.35$ were discussed in Ref.~\cite{TLS_ternary}. They are obtained from the structural relaxation time $\tau_{\alpha}$ extracted from the self-intermediate scattering function by $F_{\rm s}(q, t=\tau_{\alpha}) = 1/e$. The function is averaged over a shell of wave vectors with modulus $q = 7.2 \pm 0.05$. The onset of glassy dynamics occurs around $T_o \simeq 1.0$, signaled by $\tau_{\alpha}$ departing from the high-temperature Arrhenius behavior. The putative mode-coupling transition temperature, $T_{\rm mct}$, is obtained by fitting the relaxation times at moderate supercooling with a power law, $\tau_\alpha \propto (T-T_{\rm mct})^{-\gamma}$, resulting in $T_{\rm mct}\cong 0.62$ (with the best fit parameter $\gamma=2.7$). The lowest temperature studied is $T = 0.509$ at which the alpha-relaxation time is estimated to be four orders of magnitude larger than at $T_{\rm mct}$.

\subsection{Simulation methods}

We first generated equilibrium configurations of the model at various temperatures. This was done using the hybrid molecular dynamics/particle-swap Monte Carlo algorithm implemented in the LAMMPS package~\cite{berthier_efficient_2019}. Based on Ref. \onlinecite{parmar_ultrastable_2020} we have found that short blocks of 10 MD steps with a time step $dt=5\times10^{-3}$ interspersed with blocks of $1.75 N=21000$ attempted particle swap moves yields the most efficient dynamical speedup. At each temperature, we performed long equilibration runs ensuring the absence of aging in the dynamics and proper convergence of potential energy and relaxation time. We generated 20 independent equilibrium configurations at each temperature. 

Next, we investigated the equilibrium relaxation dynamics via standard molecular dynamics, \textit{i.e.}, without particle swaps. This was done utilizing state-of-the-art graphics-processing unit (GPU) simulations to access equilibrium dynamics at very low temperatures over unprecedented long times~\cite{rumd}. These Newtonian molecular dynamics (MD) simulations were run in the $NVT$ ensemble using a Nose–Hoover thermostat to control the temperature. The equations of motion were integrated with the timestep $dt = 0.005$. The algorithm allows one to reach a simulation time of $4.3\cdot 10^7$ ($8.6\cdot 10^9$ time steps) in $210$ hours for each of the 20 independent samples of $N = 12 000$ particles.

\subsection{Observables}

The central observable in this work is the mean-square displacement as a function of time, $\langle \Delta r^2(t) \rangle$. This quantity is defined as 

\beq
\langle \Delta r^2(t) \rangle = \frac{1}{N}\,\langle \sum_{i=1}^N |\mathbf{r}_i(t) - \mathbf{r}_i(0)|^2 \rangle~~,
\label{eq:MSD}
\eeq
in which the sharp brackets indicate an average over independent simulations and different initial times. The independent simulations were obtained either from a single very long MD trajectory or (at low temperatures) from independent initial equilibrium configurations (at higher temperatures). We present below data for the MSD both averaged over all particles and restricted to a single particle type (A, B, or C). 

At short times and low temperatures, the MSD is dominated by vibrations of the particles inside the cage formed by their neighbors. This hides a weak signal emerging from a few cage-breaking relaxation events. In order to filter out these intra-cage vibrations, we follow the procedure of Ref.~\cite{sch00,solidlike} and compute the inherent state (IS) MSD $\langle \Delta r_I^2(t) \rangle$. Recall that the inherent state is arrived at by quenching to the nearest potential-energy minimum \cite{sti83}, and the ``inherent dynamics'' is defined as the sequence of inherent states traced out by quenching all states of the ``true'' dynamics \cite{sch00a}. The inherent dynamics then defines the IS MSD, which is a continuous function of time despite the fact that the inherent dynamics itself is discontinuous. The IS MSD is computed by replacing the particle positions $\mathbf{r}_i(t)$ in Eq.~(\ref{eq:MSD}) by their inherent-state positions, $\mathbf{r}_{i, I}(t)$, which were identified by the standard conjugate-gradient method. The IS MSD computation can, of course, like the MSD itself be evaluated by averaging over all particles or restricted to a given particle type. 

\section{Results}

We investigate as mentioned the slow dynamics of the liquid by evaluating the mean-square displacement (MSD) of particles as a function of time, $\langle \Delta r^2(t) \rangle$.

\subsection{Validation of the numerical strategy}

\begin{figure}[t]
\includegraphics[width=0.48\textwidth]{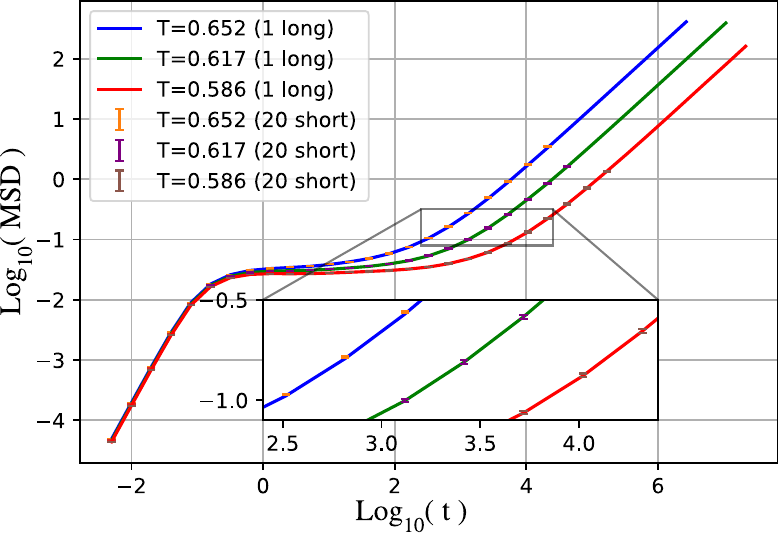}
\caption{All-particle mean-square displacement (MSD) as a function of time in log-log plots at the three highest temperatures studied. Full lines: Results of MD simulations preceded by a MD equilibration run of same length. Data points: Results averaged over 20 independent MD simulations, each initiated by an equilibrium configuration generated by particle-swap. Error bars indicate one-standard-deviation error estimates. As expected, the dynamics is the same within the error bars. 
}
\label{fig:msd}
\end{figure}

Figure~\ref{fig:msd} shows the all-particle MSD at the three highest temperatures. These data were obtained by two routes. The lines mark the MSD measured over a single long MD simulation. The points represent the data obtained by starting 20 MD simulations from independent configurations generated via the particle-swap algorithm. At all temperatures, the two curves overlap within the error bars (see insets). This confirms that equilibration by particle swap and long MD simulations yield the same results. Importantly, at temperatures lower than $T\cong 0.59$, brute-force long MD simulations can no longer reach equilibrium and particle-swap methods are necessary because of the extremely long MD equilibration times.

\begin{figure*}[t]
\includegraphics[width=\textwidth]{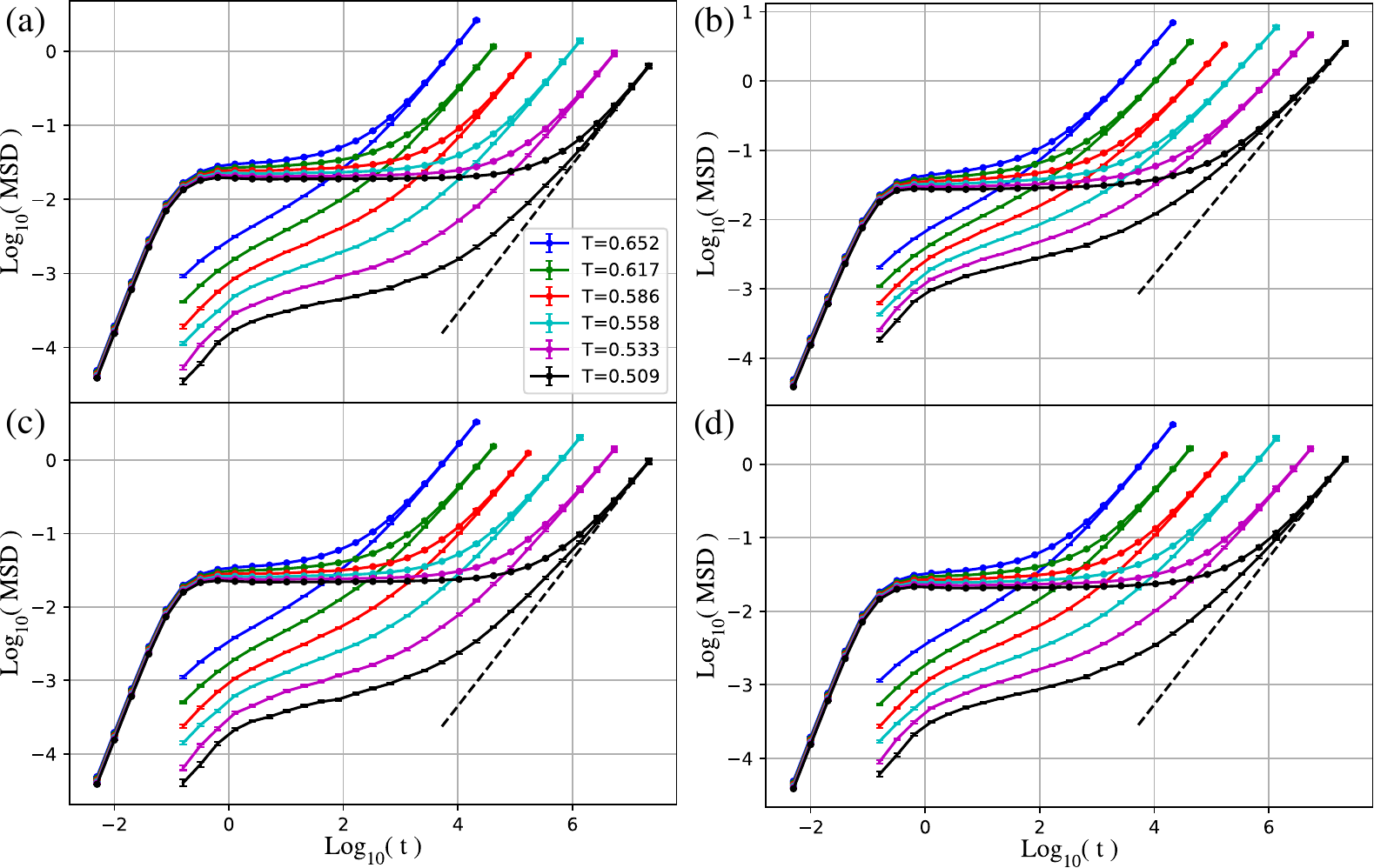}
\caption{Mean-square displacement as a function of time in log-log plots for (a) A, (b) B, (c) C, and (d) all particles at all temperatures studied. All initial configurations were equilibrated by swap. For each temperature, two curves are shown: The thermal MSD (for the three highest temperature these are the same data as in Fig.~\ref{fig:msd}), and the inherent dynamics. 
}
\label{fig:imsd} 
\end{figure*}

\subsection{Thermal and inherent MSD results}

Figure~\ref{fig:imsd} shows the MSD at six temperatures, five of which are below $T_{\rm mct}\cong 0.62$. The figure shows data averaged over each particle type (A, B, or C) separately, as well as over all particles. All temperatures investigated are in the viscous regime and the MSD curves exhibit the same qualitative behavior: The dynamics is ballistic at very short times before the particles collide with their nearest neighbors, resulting in a transient caging. From the microscopic time scale $t\simeq1$ and onward (in simulation units), the MSD reaches a plateau reflecting that the main motion is of a vibrational nature. The plateau length increases with decreasing temperature and reaches five decades at the lowest temperature investigated. At long times, the MSD data cross over to diffusive behavior, $\langle \Delta r^2(t) \rangle \propto 6Dt$, in which $D$ is the (temperature-dependent) diffusion coefficient. 

As previously reported~\cite{solidlike,30ms}, the IS MSD is non-zero and increases with time also in the MSD plateau regime. This reflects the fact that structural rearrangements occur already at short times and that these accumulate over time. At long times, all particles have undergone structural rearrangements and the thermal and inherent MSDs coincide.

\subsection{Fitting the inherent MSD}

Next, we focus on the IS MSD in order to investigate the validity and predictability of two fitting forms, the von Schweidler law and the random barrier model. 

\begin{figure*}[t]
\includegraphics[width=\textwidth]{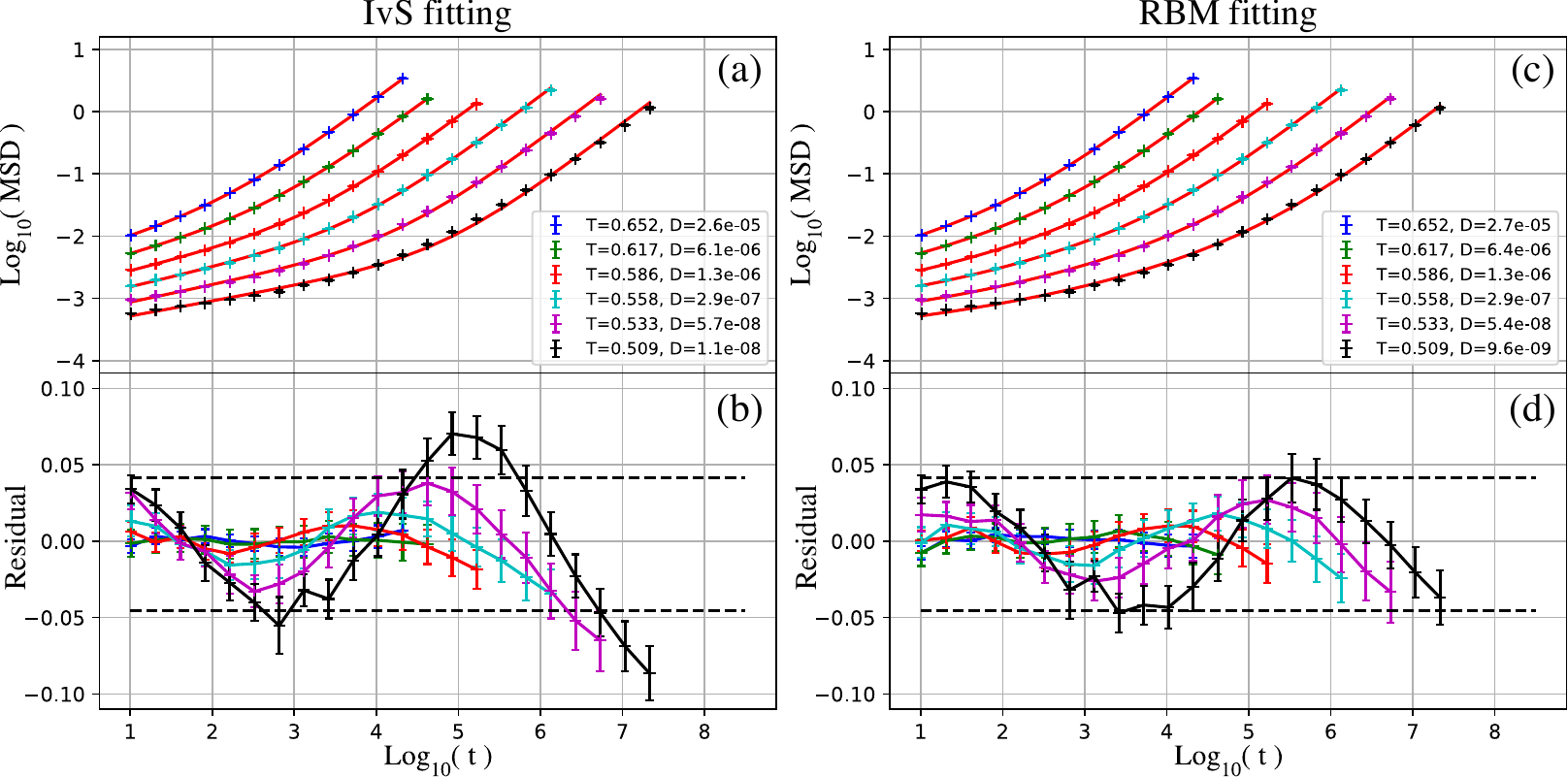}
\caption{ 
Fits (a,c) and residuals (b, d) for the inherent all-particle MSD data (crosses). (a, b)) Fitting to the inherent von Schweidler (IvS, Eq. (\ref{eq:vonschweidler})). (c, d) Fitting to the Random Barrier Model (RBM, Eq. (\ref{eq:FitRBMs})). The residuals are computed as $\log_{10}(\langle \Delta r^2(t) \rangle  / \textrm{MSD}_{fit})$. Between the black dashed lines in (b, d), the fits are less than 10\% off from the data. Despite having one less parameter, the RBM model fits the low temperature data better than the inherent von Schweidler law. In particular, the RBM fits the simulation data significantly better at long times. 
}
\label{fig:imsd_fits_All}
\end{figure*}

\begin{figure*}[t]
\includegraphics[width=\textwidth]{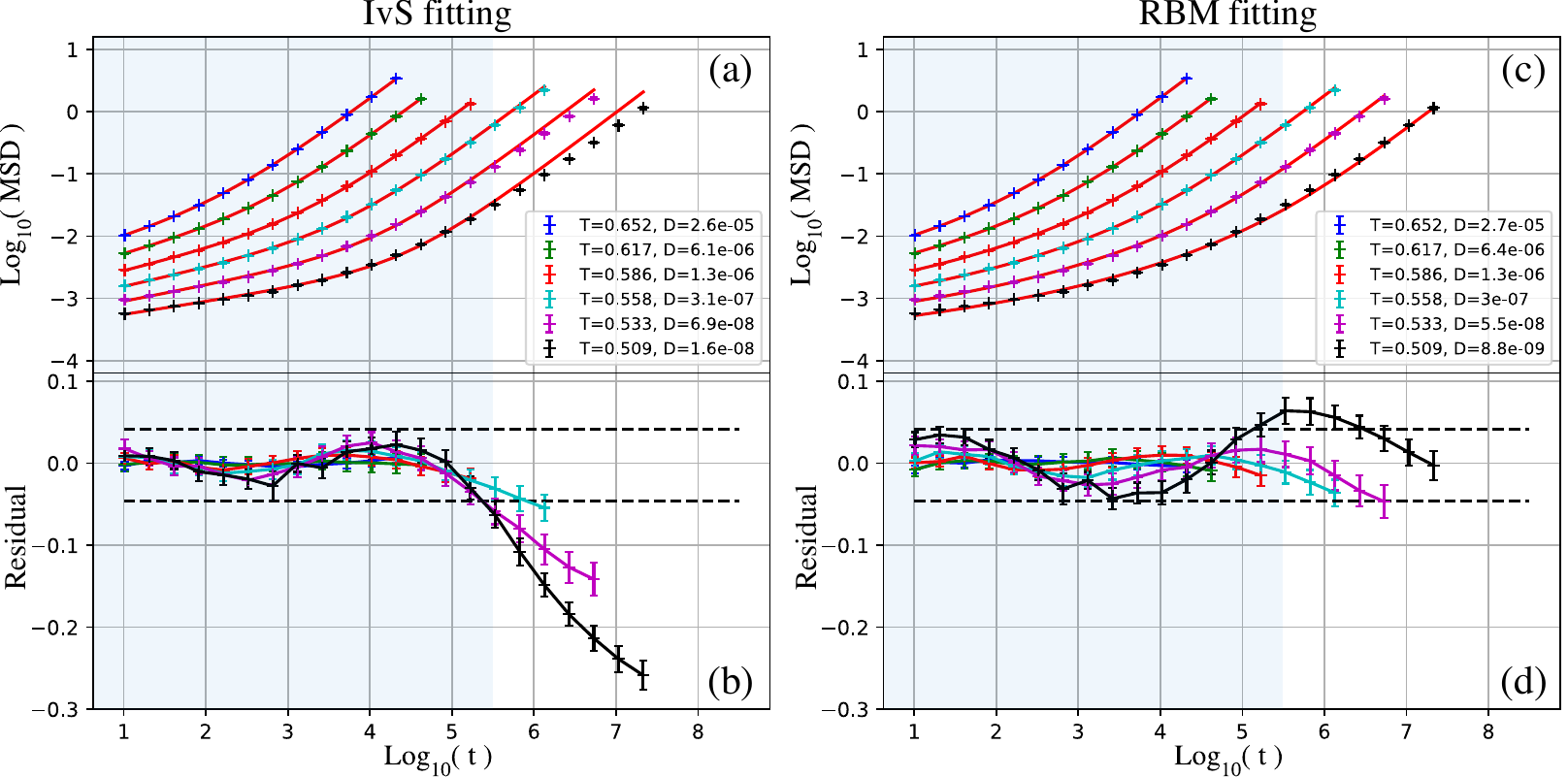}
\caption{\label{fig:imsd_restricted_fits_All} 
Same as \fig{fig:imsd_fits_All}, except that fits are here restricted to $t<3\cdot 10^{5}$, highlighted by the shaded region. 
}
\label{fig:imsd_fits_All_restricted}
\end{figure*}

\subsubsection{von Schweidler fit}

MSD simulation data are conveniently fitted to the von Schweidler empirical expression, which emerges from MCT theory~\cite{kob_testing_1995, solidlike}

\beq
\Delta_{IvS}(t)= r_0^2 + a(6Dt)^b + 6Dt\,,
\label{eq:truevonschweidler}
\eeq
in which $b$ is a non-universal exponent. Since our focus is on the inherent dynamics that removes the influence of vibrations leading to the plateau in the MSD, we define an ``inherent'' von Schweidler (IvS) law by setting $r_0^2 = 0$ in Eq.~(\ref{eq:truevonschweidler}):
\beq
\Delta_{IvS}(t)= a(6Dt)^b + 6Dt\,.
\label{eq:vonschweidler}
\eeq
Equation~(\ref{eq:vonschweidler}) can be expressed as $l_0^2\left[\left(t/\tau_0\right)^b + t/\tau_0\right]$ in which $l_0^2 \equiv a^{1/(1-b)}$ and $\tau_0 \equiv l_0^2/(6D)$. Thus Eq. (\ref{eq:vonschweidler}) has one dimensionless shape parameter, $b$.

We show in Fig.~\ref{fig:imsd_fits_All}(a) the all-particle inherent MSD curves and their fits to Eq.~(\ref{eq:vonschweidler}) (data for the A, B, and C particles separately are discussed in the Appendix). We avoid the very short time ballistic regime by focusing on times $t > 10 $ (LJ units), corresponding to the MSD plateau and diffusive regimes. 

At all temperatures and times, the fit appears to be quite good. We quantify the quality of the fit by computing the residuals defined as the (base 10) logarithm of the ratio of the IS MSD to the von Schweidler fit, see Fig.~\ref{fig:imsd_fits_All}(b). When the residuals are positive (resp. negative), the von Schweidler fit underestimates (resp. overestimates) the IS MSD. Within the region delimited by the two dashed lines, the logarithm of the fit is less than 10\% off from the data. The quality of the fit deteriorates with decreasing temperature. This may be explained by the fact that Eq. (\ref{eq:vonschweidler}) was proposed in the context of MCT, the approximations of which break down for the five temperatures investigated below $T_{\rm mct}$.

\subsubsection{Random Barrier Model fit}

We now turn to the fit to the Random Barrier Model (RBM) prediction. The RBM is a model for the motion of particles in a random environment, characterized by the following simplifying assumptions \cite{dyr85,dyr88,sch00}: 
\begin{enumerate}
\item Particles jump stochastically on a cubic lattice in $d$ dimensions;
\item Only nearest-neighbor jumps are possible;
\item All site energies are identical;
\item The particles do not interact, i.e., even self-exclusion at each lattice site is ignored;
\item The link jump rates are proportional to $\exp(-\beta E)$ in which $\beta=1/k_BT$ and $E$ is an activation energy chosen randomly from a probability distribution, $p(E)$.
\end{enumerate}
The RBM is characterized by a universal time-dependent MSD in the extreme disorder limit. This limit is defined by $\beta\to\infty$, i.e., when jump rates cover several decades. This means that $\langle \Delta r^2(t) \rangle$ in dimensionless units becomes independent of $p(E)$ in the extreme disorder limit. This universality, which has been validated by extensive simulations \cite{sch00,sch08}, follows from the fact that percolation controls the physics in this limit \cite{sch08} -- the only assumption needed is that $p(E)$ is continuous at the energy at the (link) percolation threshold \cite{sch00,sch00c}.

The frequency-dependent diffusion coefficient is defined by 
\beq
D(\omega)\equiv (-\omega^2/6)\int_{0}^{\infty}\langle\Delta r^2(t)\rangle\exp(i\omega t)dt\,,
\eeq
see Refs. \cite{hau87,bou90,dyr02}. An accurate analytical approximation to the RBM universal $D(\omega)$ is the solution of $\ln\tD=(i\tom/\tD)^{2/3}$ in which $\tom$ is a scaled frequency and $\tD\equiv D(\omega)/D(0)$ \cite{sch08}. This is derived by combining the Alexander-Orbach conjecture that the percolation cluster has harmonic dimension $4/3$ \cite{ale82} with the effective-medium approximation (EMA) applied to diffusion on the percolation cluster defined by the smallest activation-energy links \cite{sch08}. Even though both the EMA and the Alexander-Orbach conjecture are known not to be rigorously correct in all dimensions, the quoted equation provides an excellent fit to computer simulations of the RBM in three dimensions \cite{sch08}. Of particular interest in the present context is that the Alexander-Orbach conjecture is independent of the nature and dimension of the space in which the percolation takes place (the Alexander-Orbach conjecture for the harmonic dimension of the percolation cluster, 4/3, applies rigorously above the upper critical dimension, which is six \cite{hug96}), but is surprisingly accurate at lower dimensions. A transition to a frequency-independent diffusion coefficient takes place at low frequencies, corresponding to the long-time $\langle\Delta r^2(t)\rangle\propto Dt$ behavior. In this region, the RBM analytical prediction needs an extra term and is well described by the solution of the following empirical equation $\ln\tD=(i\tom/\tD)(1+(8/3)i\tom/\tD)^{-1/3}$ \cite{sch08} (note that the scaled frequencies in the two analytical RBM expressions are proportional, but not identical). The Appendix quotes a numerical approximation to the RBM mean-square displacement as a function of time that is accurate over ten decades of time.

We report in Fig.~\ref{fig:imsd_fits_All}(c) the IS MSD along with their best fit to the RBM prediction. To assess the validity of the RBM prediction, we report in Fig.~\ref{fig:imsd_fits_All}(d) the residuals defined as above. Despite having no shape parameters -- as compared to the one shape parameter of the von Schweidler expression -- the RBM describes the data better over the range of temperatures investigated. While the two analytical approximations work equally well close to $T_{\rm mct} \sim 0.62$, the error at the lowest temperatures is clearly smaller for the RBM prediction. 

\begin{figure}[t]
\includegraphics[width=0.48\textwidth]{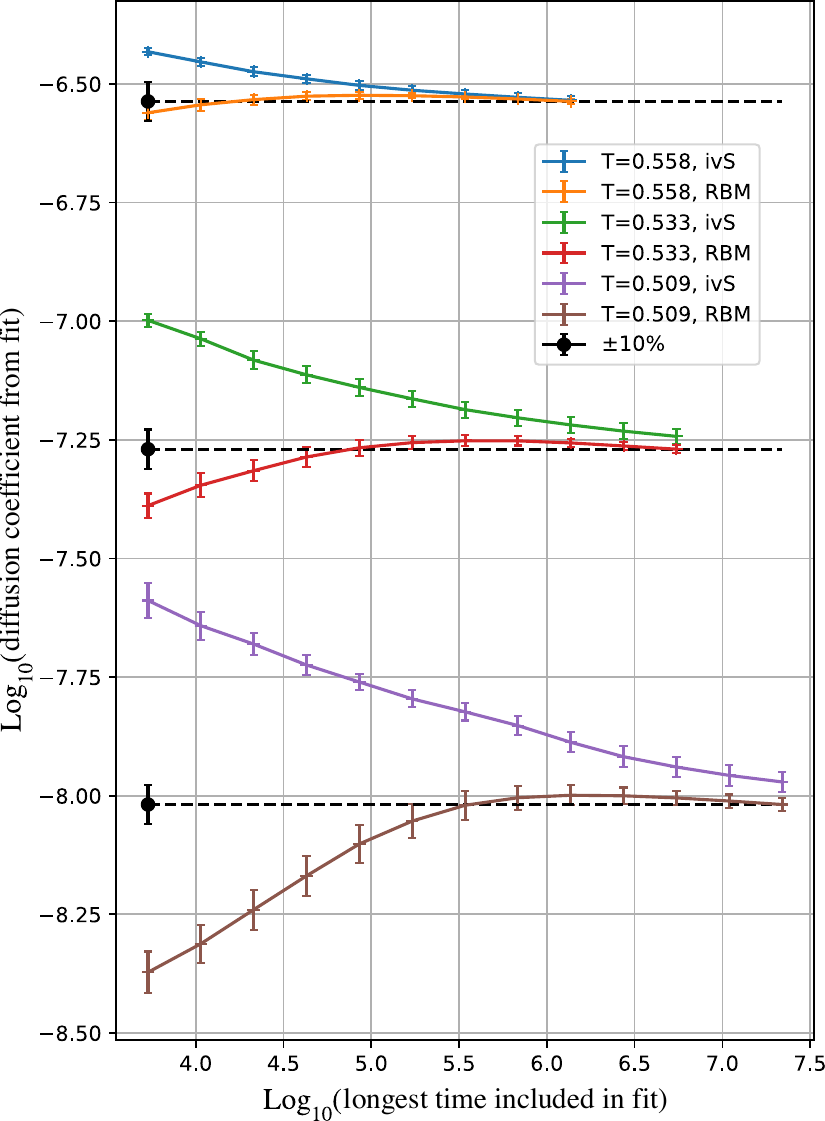}
\caption{
Diffusion constant obtained by fitting the von Schweidler and RBM predictions to the IS MSD over a time window starting at 10 LJ time units, terminating at the time indicated on the horizontal axis. The horizontal dashed lines highlight the value obtained for fitting the whole trajectory, with the bullet/error bar (left) indicating an error of 10\% around it. Temperature decreases from top to bottom. 
}
\label{fig:D_fits_All}
\end{figure}

\begin{figure}[t]
\includegraphics[width=0.48\textwidth]{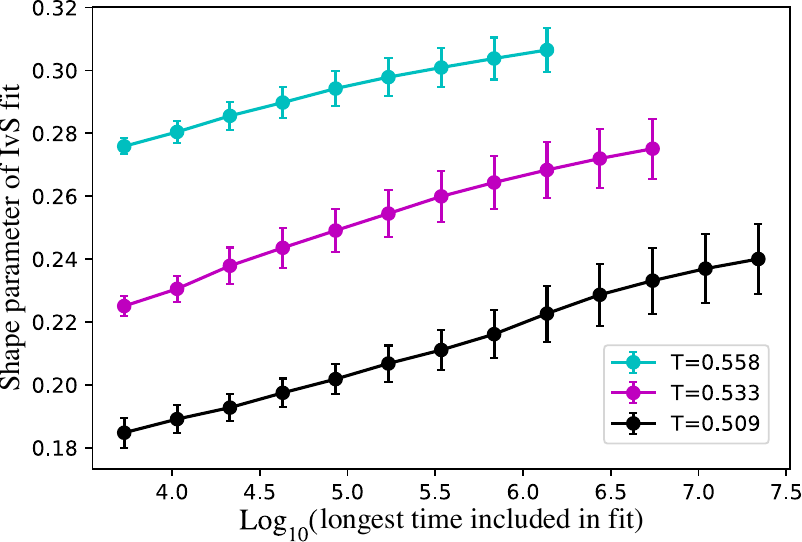}
\caption{The shape parameter, $b$, of the von Schweidler fit to the IS MSD as a function of the maximum time included in the fit.
}
\label{fig:a_fits_All} 
\end{figure}

\subsection{Predictive power of the two fits}

Given that long simulations are computationally costly, one might find it useful to do extrapolations of the data at long times and low temperature. We focus on the IS MSD and use our long simulation trajectories to investigate how well the two fitting functions predict the IS MSD at long times. We do this by restricting the fits to shorter times and then comparing the long-time data predicted from the fit to the measured one. The fits were performed over a window $10 < t < 3\cdot 10^5$, i.e., two orders of magnitude shorter than the longest simulation. Data, restricted fits, and their residuals are shown in Fig.~\ref{fig:imsd_fits_All_restricted}. We see in Fig.~\ref{fig:imsd_fits_All_restricted}(a) that the restricted von Schweidler fit significantly departs from the data at long times. This is confirmed by the residuals (b), which plummet at long times and low temperature. The RBM extrapolation is much more accurate, see Fig.~\ref{fig:imsd_fits_All_restricted}(c,d). 

The above extrapolation yields a diffusion coefficient, even when the diffusive regime cannot be accessed brute-force by long simulations. This may be useful to test Stokes-Einstein breakdown in the extremely viscous regime, if combined with, e.g.,  a parabolic law to extrapolate the relaxation time data. We investigate stability and convergence of the estimated diffusion coefficient by restricting the fits over time windows of increasing length. The estimated diffusion coefficient is reported in Fig.~\ref{fig:D_fits_All} as a function of the longest time included in the fit (the shortest one still being 10 LJ time unit) for both the von Schweidler and RBM fits. Fitting the data over too short time scales yields bad estimates in both cases. As anticipated in Fig.~\ref{fig:imsd_fits_All_restricted}, the IvS fit overestimates the diffusion coefficient. The estimate obviously improves with increasing the fitting time windows. The convergence is much faster for the RBM fit compared to that of the IvS, however, especially at low temperature where the long simulation times are not enough to observe convergence of the IvS estimate. The RBM prediction yields a diffusion coefficient, which varies less than 10\% when the fitting window increases by two decades. The fact that the parameter fit, the diffusion coefficient for the RBM prediction, is not very sensitive to the fitting range confirms that it represents well the data over a broad range of temperatures and time scales. 

We now turn to the evolution of the IvS shape parameter, $b$, with the fitting range. The results are shown in Fig.~\ref{fig:a_fits_All} at three temperatures, decreasing from top to bottom. The fitting starts at 10 LJ time units and terminates at the value indicated on the horizontal axis. We see that the shape parameter varies significantly with the fitting window. At the lowest temperature, the shape parameter increases by more than 30\% when increasing the fitting window by three orders of magnitude, and the shape parameter does not converge over a simulation window of nine decades, the maximum accessible. We interpret this as reflecting an inherent weakness of the IvS fit, which is based on assuming a power law in time.

\section{Discussion}

The RBM prediction accurately captures the time evolution of the IS MSD down to unprecedentedly low temperatures, extending the findings of Ref.~\cite{solidlike} to a regime closer to the experimental glass transition. We have shown that the RBM outperforms the von Schweidler law in its ability to extrapolate the data to longer times and reliably extract the diffusion coefficient. This is striking in view of the fact that the RBM has no shape parameters, in contrast to the von Schweidler law, which has one shape parameter.

\begin{figure*}[t]
\includegraphics[width=\textwidth]{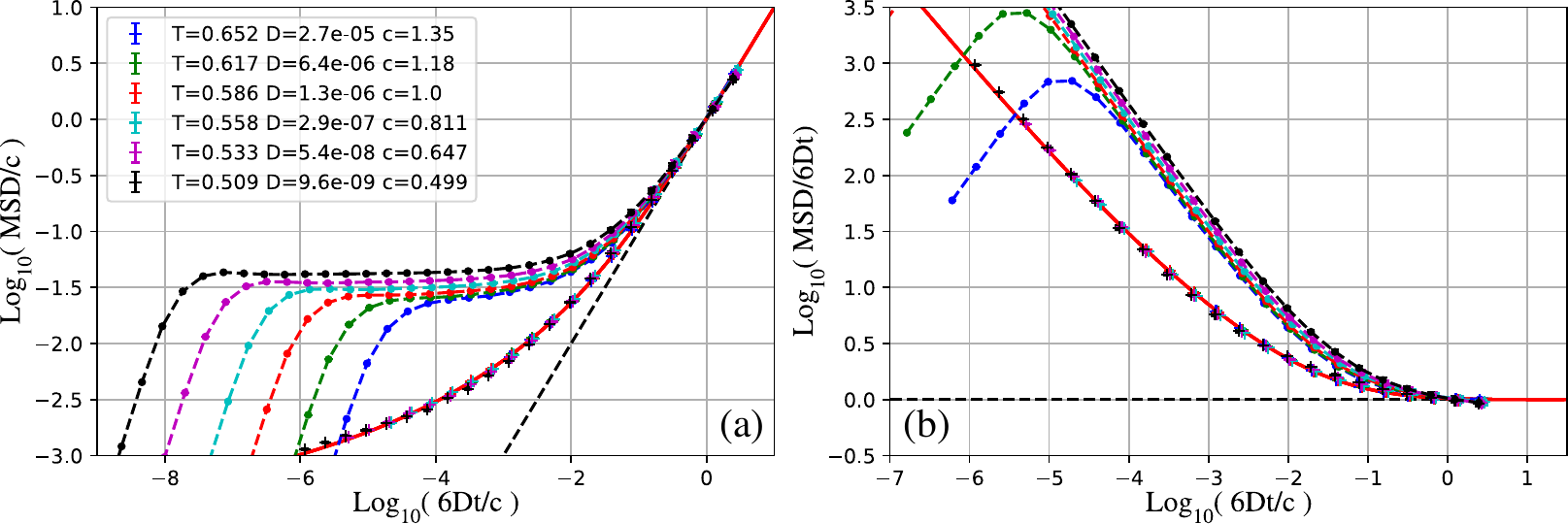}
\caption{\label{fig:imsd_scaled_All} 
(a) True and inherent all-particle MSD scaled by $D$ and $c$ as estimated by RBM fits. Red full curve: RBM prediction of the inherent MSD.  (b) Same data divided by $6Dt$.
}
\end{figure*}

What do our results tell about the possible universality discussed in the introduction? We have shown that the IS MSD is well reproduced by the RBM prediction. Once the IS MSD data are rescaled by a parameter $c$ with dimension of a length squared and time by $c/(6D)$, the curves collapse for all temperatures investigated, see Fig.~\ref{fig:imsd_scaled_All} (plus-symbols). Corresponding plots for A, B, and C particles separately are given in the Appendix. This implies that time-temperature superposition applies to the IS MSD and, importantly, that the IS MSD of the studied model adheres to the possible universality discussed in the introduction. 

Figure~\ref{fig:imsd_scaled_All} shows also the true/thermal MSD data (connected by dashed lines), scaled with the parameters found from the IS MSD. In contrast to the binary model studied in Ref.~\cite{solidlike}, the true/thermal MSD does \textit{not} collapse onto a master curve by this scaling procedure. This supports the conjecture that the additive constant needed to model the intermediate-time cage rattling is non-universal \cite{solidlike}, which might explain why the apparent RBM universality has not been observed earlier. 

\begin{figure*}[t]
\includegraphics[width=\textwidth]{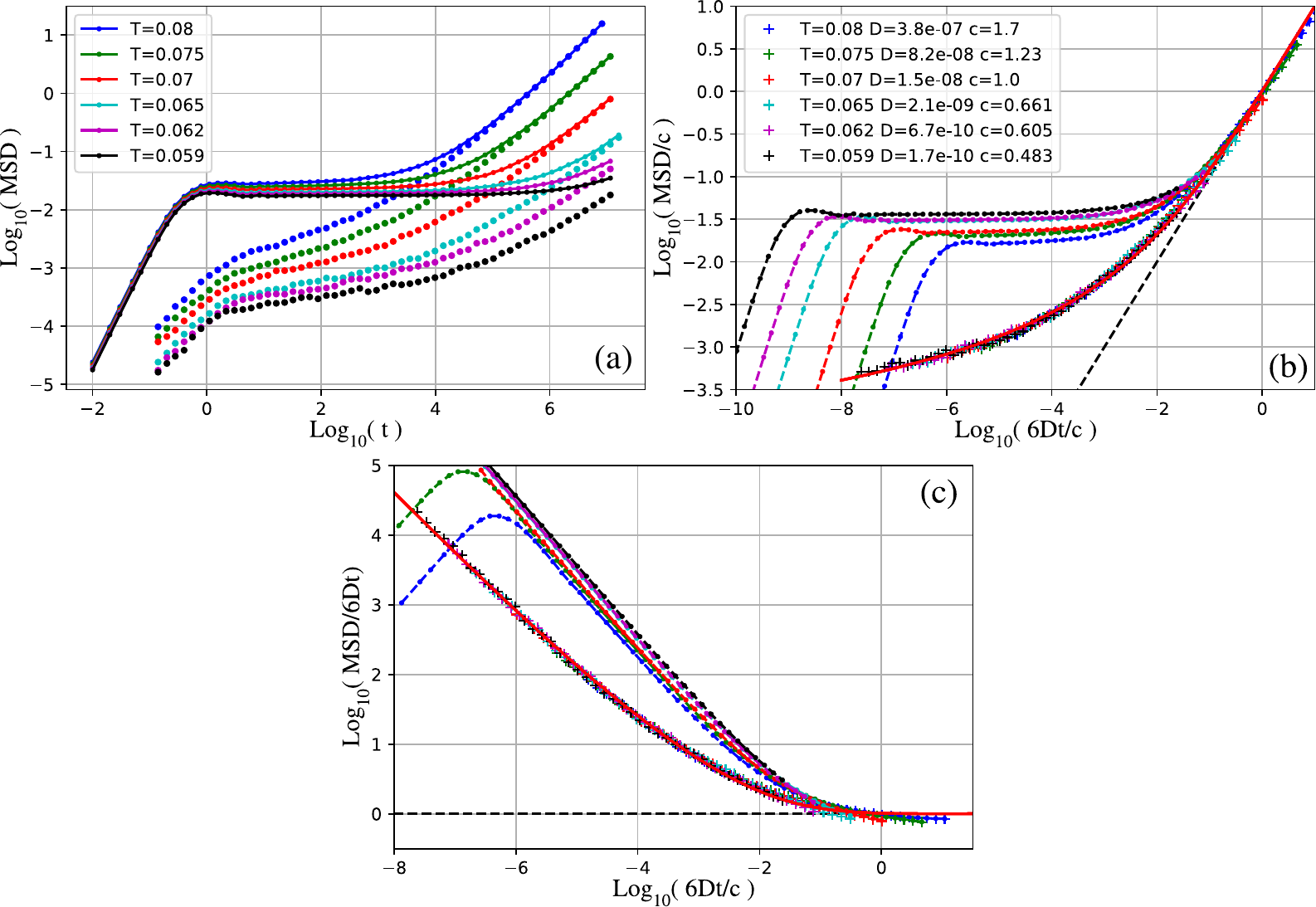}
\caption{\label{fig:imsd_scaled_polydisperse} 
(a) True and inherent all-particle MSD for the polydisperse model. (b) Same data, scaled by $D$ and $c$ as estimated by RBM fits. Inherent data only used for $t>10$. Red full curve: RBM prediction of the inherent MSD.  (c) Same as (b), by y-axis divided by $6Dt$.
}
\end{figure*}

Finally, we present selected data on a third, polydisperse LJ model, whose structural relaxation dynamics has been analyzed down to the extrapolated experimental glass transition in Ref. \cite{30ms}. We have performed the RBM analysis for the IS MSD of this additional model. Figure~\ref{fig:imsd_scaled_polydisperse} shows the thermal and IS data (a), and once rescaled by $D$ and $c$ (b). While the present IS MSD data for the polydisperse model is more noisy, we again find an overall good agreement with the RBM for all temperatures studied. This significantly increases the regime of validity of the RBM scaling, since the lowest temperature studied $T = 0.059$ has an extrapolated relaxation time which is 11 decades longer than at the onset temperature of glassy dynamics, i.e., almost at the experimental glass transition temperature typically defined by a 12-decades slowdown (this is almost seven decades slower than MCT). The polydisperse model also reveals that the intermediate-time additive constant is non-universal, see Figure~\ref{fig:imsd_scaled_polydisperse}(b).

\begin{figure*}[t]
\includegraphics[width=\textwidth]{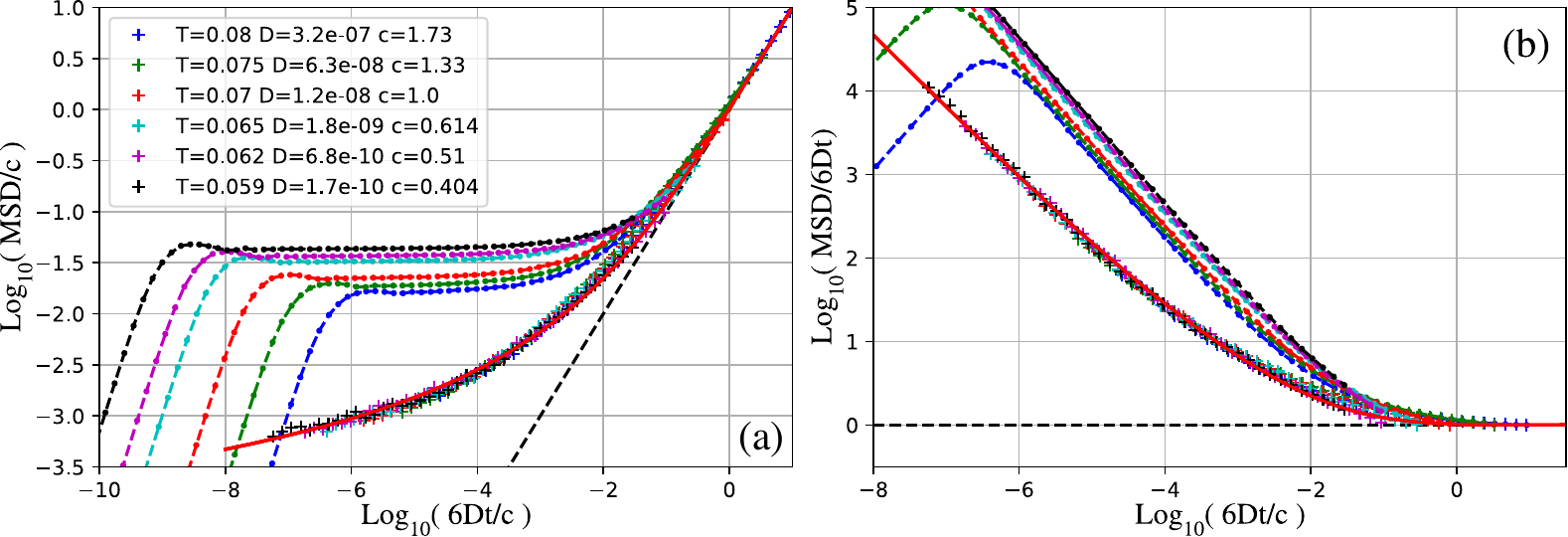}
\caption{\label{fig:imsd_scaled_polydisperse_minD} 
Same data as in Fig.~\ref{fig:imsd_scaled_polydisperse}, but the RBM fit was performed with the constraint on the diffusion coefficient that the MSD for the last point for each temperature is \emph{not} not allowed to be smaller than $6Dt$.
}
\end{figure*}

We note that in Fig.~\ref{fig:imsd_scaled_polydisperse}(c) it is apparent that the RBM estimate of the diffusion coefficient is about 0.1 decade (corresponding to roughly 25\%) too big when compared to the long-time limit of the simulation data. In comparison, the diffusion coefficient changes by more than a factor of 1000 over the temperature interval studied. Looking closely at Fig.~\ref{fig:imsd_scaled_All}(b), we see that extrapolating the data by eye suggests a similar behavior for the ternary model. Exactly where the small deviances are found depends how the RBM fit is performed; For the data in Fig.~\ref{fig:imsd_scaled_All} and Fig.~\ref{fig:imsd_scaled_polydisperse} the RBM fit was used to estimate the two scaling parameters $D$ and $c$ simultaneously. In contrast, for the binary system in Ref.~\cite{solidlike} where all state points had a well-defined diffusive regime, the diffusion coefficient was first estimated by fitting the long time MSD, and only after this was $c$ estimated by fitting to the RBM prediction. Unsurprisingly, this procedure gives estimates of the diffusion coefficient consistent with the long-time limit of the simulation data. We can not apply this procedure to the polydisperse system, since the diffusional regime can only be reached for a few of the state points. 
What we can do instead, is to put a constraint on the fitted diffusion coefficient so that the MSD for the last data point for each temperature is \emph{not} smaller than $6Dt$. The result of this procedure is shown in Fig.~\ref{fig:imsd_scaled_polydisperse_minD}, now showing good consistency between the  diffusion coefficients and the long-time limit of the simulation data.   

\begin{figure*}[t]
\includegraphics[width=\textwidth]{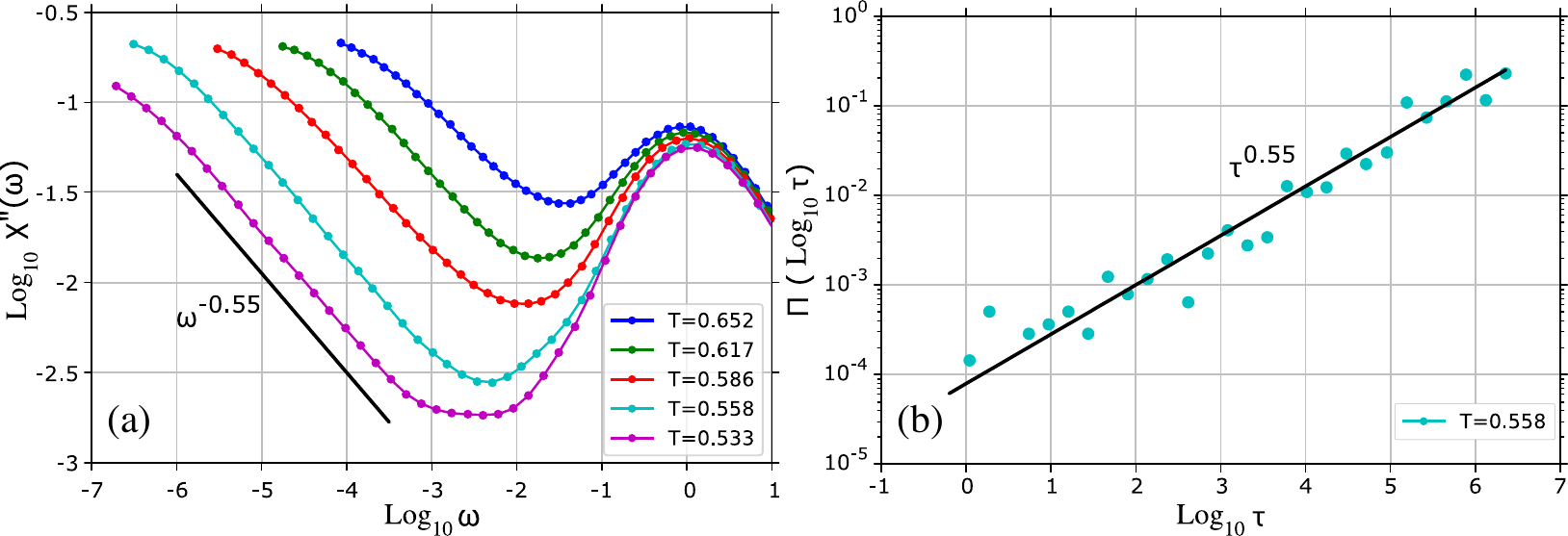}
\caption{\label{fig:relaxation} 
Relaxation spectra and apparition of mobile regions in the ternary model. (a) Relaxation spectra $\chi ''(\omega)$ as a function of frequency $\omega$, defined from the self-intermediate scattering function. The data reveals an ``excess wing'', as observed in the polydisperse model, yet with a different exponent, here 0.55 (0.38 for the polydisperse model). (b) Probability distribution $\Pi$ of the time $\tau$ at which new mobile regions appear in the liquid. The distribution is characterized by the same exponent 0.55. 
}
\end{figure*}

In summary, three different model liquids show very good -- though not perfect -- agreement between the inherent MSD and the RBM prediction, which is remarkable in view of the fact that the RBM has no shape parameters in the extreme disorder limit. It might be tempting to conclude that these models are very similar; after all, they are atomic Lennard-Jones liquids with two, three, and many types of particles. Fig.~\ref{fig:relaxation} shows, however, that in regard to relaxation, the ternary model is quantitatively different from the polydisperse model. We reproduce the analysis performed in Ref. ~\cite{30ms} on the polydisperse model for the ternary one: we compute the relaxation spectra $\chi ''(\omega)$ defined from the self-intermediate scattering function. We find that the relaxation spectra of the ternary model also exhibits a power-law scaling at low frequency. The exponent found here, 0.55, is however different from that found in the polydisperse model (0.38). We confirm that the power-law in the spectra arises from the apparition of mobile regions, which occurs over very broadly distributed timescales. Indeed, Fig.~\ref{fig:relaxation}(b) shows that the probability distribution of the time $\tau$ at which new mobile regions appear exhibits a power-law with the same exponent. We conclude that the relaxation process is qualitatively similar, yet quantitatively different, in the two models: mobility starts in rare regions over broad timescales, which are known to then slowly grow in time. This observation is interesting because the short-time signal in the IS MSD --well described by the RBM in the two models-- is directly caused by the apparition of these rare mobile regions, yet characterized by different exponents. 

More work is needed to determine whether the RBM is a truly universal representation of the inherent MSD of very viscous glass-forming liquids. In particular, it would be interesting to investigate molecular models, which has recently become possible by swap methods \cite{boh25a,sim25}. It is imperative to explain why a model assuming sites of identical energy, i.e., a zero ``inherent'' specific heat, can represent data for the dynamics of realistic models of glass-forming liquids. This constitutes an important challenge to future theory development.

\begin{acknowledgments}
This work was supported by the VILLUM Foundation's \textit{Matter} grant (VIL16515). The authors thank Ludovic Berthier for helpful suggestions and discussions. We thank Benjamin Guiselin for the IS MSD of the three-dimensional polydisperse model.
\end{acknowledgments}

  \section{Appendix}


The Random Barrier Model (RBM) was solved numerically in the real-Laplace-frequency domain on a cubic lattice with a box distribution of energy barriers \cite{sch08}. At low temperatures the shape of the frequency-dependent diffusion coefficient $D(s)$ for the RBM becomes universal, i.e., independent of temperature and energy barrier distribution \cite{sch00a}. $D(s)$ for $\beta =320$ where $\beta$ is the maximum barrier over the temperature, was found to be a good representation of the universal RBM prediction.

To facilitate transformation from the real-Laplace-frequency domain to the time domain over the many decades involved, we fit to the following empirical fitting function in which $\tilde s \equiv s/D(s)$
 
\begin{equation}
  \frac{D(\tilde s)}{D_0} = 1 + \sum_{j=1}^{10} a_j\tilde s^{j/10} \label{eq:FitRBMs}
\end{equation}
The fitting was performed after taking the logarithm of both the numerical data and the fitting function, resulting in the following parameters $(a_1, ..., a_{10})$ =
(-1.1914,  11.2368, -34.2903, 26.6019, 47.0002, -96.2905, 77.4671, -27.0535, 7.72535, -0.178844). 

Using $D(s)\equiv (s^2/6)\int_{0}^{\infty}\langle\Delta r^2(t)\rangle\exp(-st)dt$ (see the main text), Eq. (\ref{eq:FitRBMs}) implies

\begin{equation}
  \langle\Delta r^2(\tilde t)\rangle = 6D_0\tilde t \left(1 + \sum_{j=1}^{10} \frac{a_j {\tilde t}^{-j/10}}{\Gamma(2-j/10)}\right) \label{eq:Fit}
\end{equation}
This is the equation used to represent the shape of the RBM in this paper. \Eq{eq:Fit} corrects a typo in Eq. (A2) of Ref. \onlinecite{solidlike} (in which ${\tilde t}^{-j/10}$ erroneously appeared as ${\tilde t}^{1-j/10}$), a typo that was also corrected in an Erratum at the time.

\begin{figure*}[t]
\includegraphics[width=0.32\textwidth]{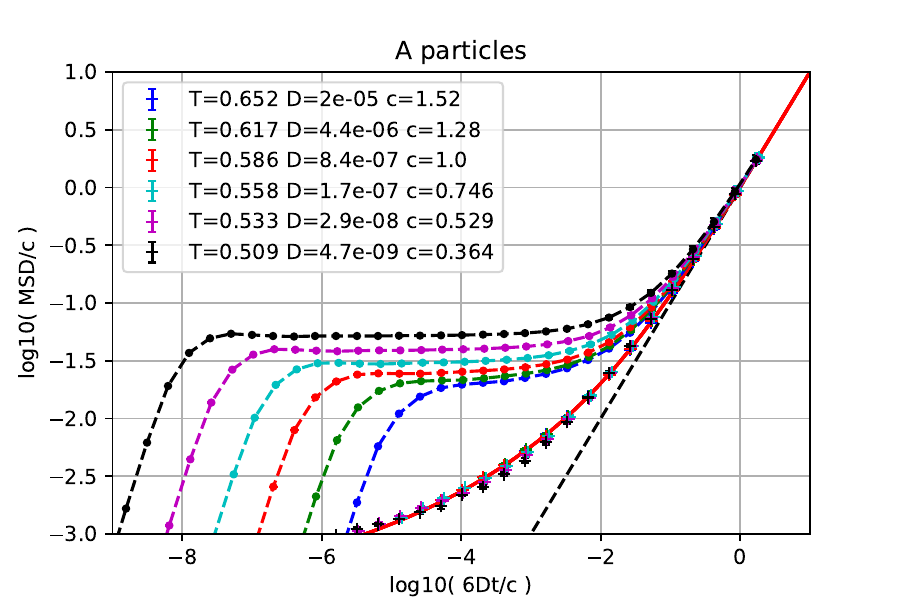}
\includegraphics[width=0.32\textwidth]{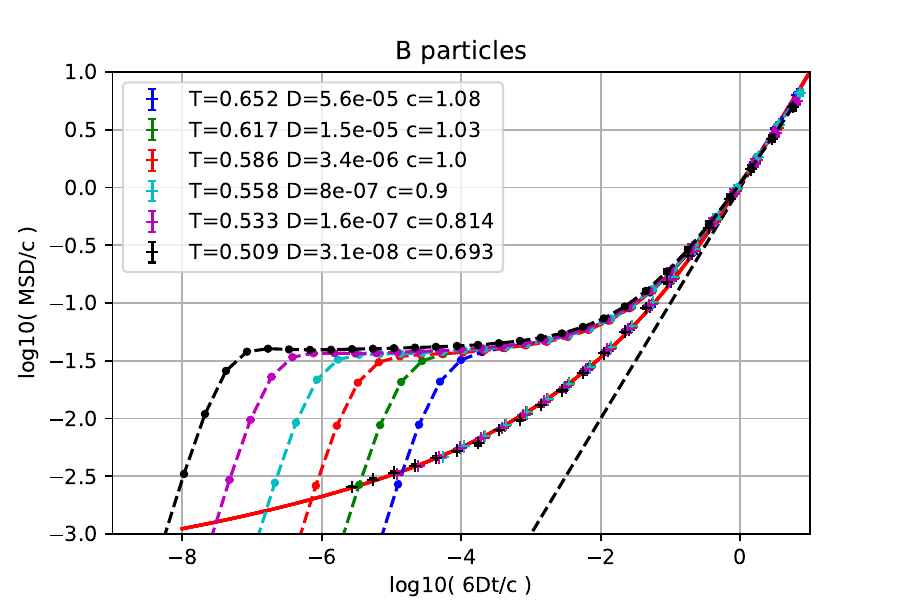}
\includegraphics[width=0.32\textwidth]{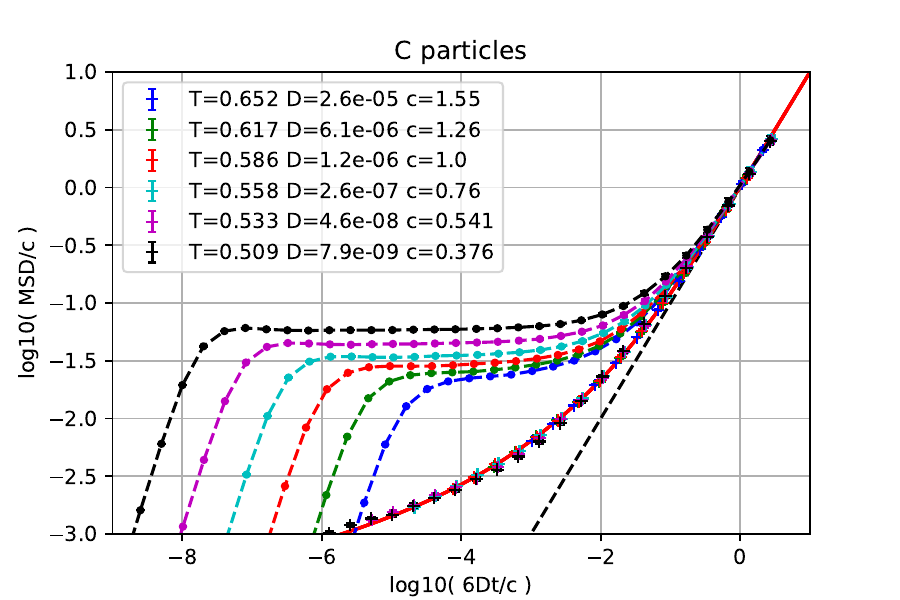}\\
\includegraphics[width=0.32\textwidth]{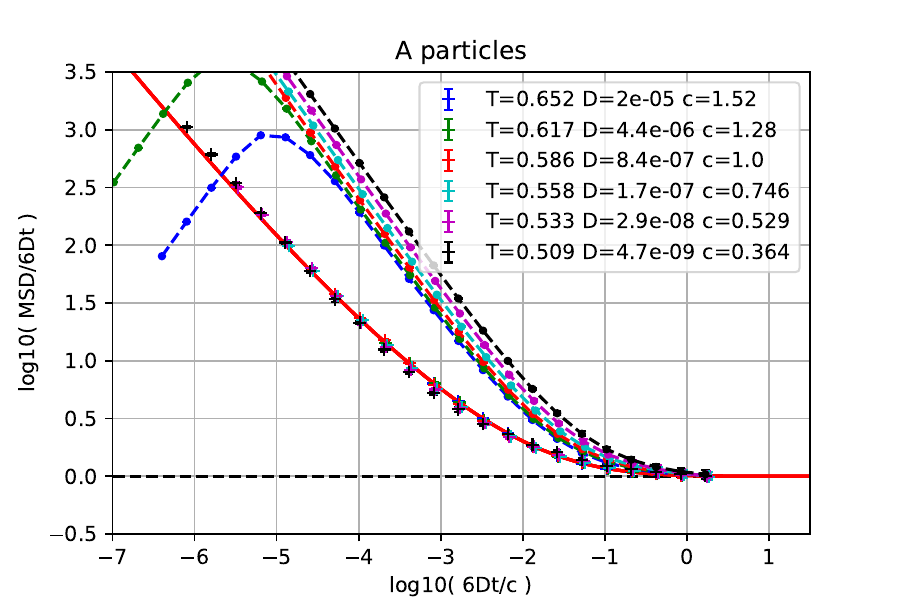}
\includegraphics[width=0.32\textwidth]{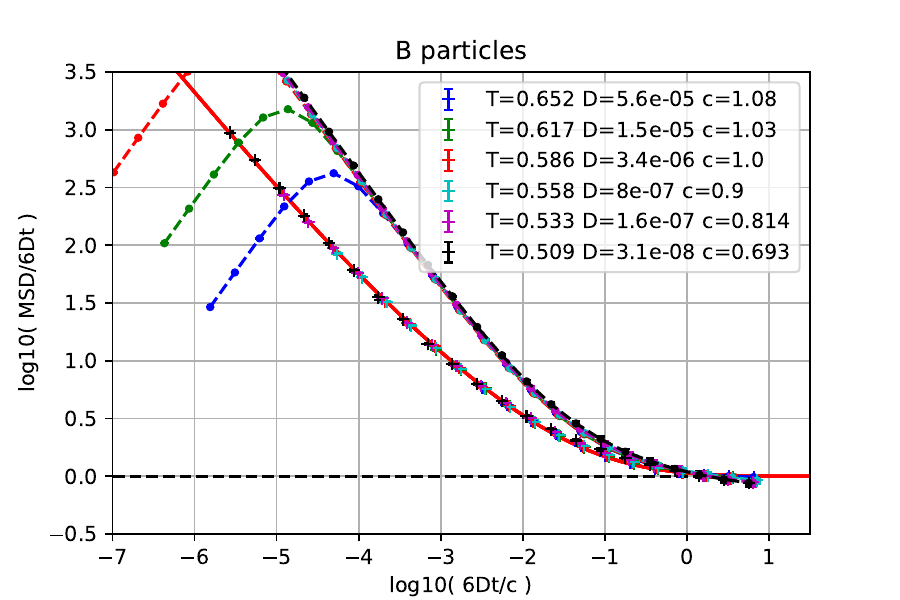}
\includegraphics[width=0.32\textwidth]{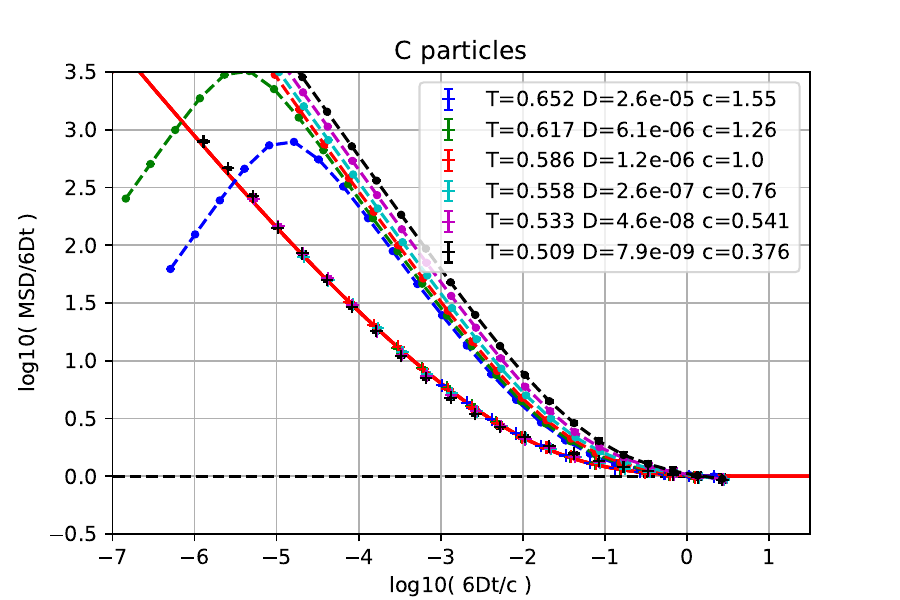}
\caption{\label{fig:imsd_scaled_A} 
Upper panels: True and inherent A-particle MSD scaled by D and c as estimated by RBM fits for A, B, and C particles respectively. Red full curve: RBM prediction of the inherent MSD. Lower panels: Same data divided by $6Dt$.
}
\end{figure*}

\bibliography{main.bib,jcd}

\end{document}